\documentclass[reprint,showpacs,superscriptaddress,amsmath,amssymb,aps,prb]{revtex4-1}

\usepackage{graphicx}
\usepackage{bm}
\usepackage{multirow}
\usepackage{color}
\usepackage{dcolumn}
\usepackage{hyperref}
\hypersetup{colorlinks=true,citecolor=blue,linkcolor=blue,urlcolor=blue,anchorcolor=blue}
\usepackage{xspace}

\newcommand{\bcgo}{\texorpdfstring{Ba\textsubscript{2}CoGe\textsubscript{2}O\textsubscript{7}}{Ba2CoGe2O7}\xspace}       
\newcommand{\ccso}{\texorpdfstring{Ca\textsubscript{2}CoSi\textsubscript{2}O\textsubscript{7}}{Ca2CoSi2O7}\xspace}       
\newcommand{\bmgo}{\texorpdfstring{Ba\textsubscript{2}MnGe\textsubscript{2}O\textsubscript{7}}{Ba2MnGe2O7}\xspace}       
\newcommand{\bcugo}{\texorpdfstring{Ba\textsubscript{2}CuGe\textsubscript{2}O\textsubscript{7}}{Ba2CoGe2O7}\xspace}       

\newcommand{\PFTOm}{$P\bar{4}2_1m$\xspace}           
\newcommand{\TN}{$T_\mathrm{N}$\xspace}              
\newcommand{\degree}{$^\circ$\xspace}                

\usepackage{physics}
\usepackage[dvipsnames]{xcolor}
\definecolor{newContentColor}{rgb}{0.00,0.0,0.75}

\newcommand{\nc}[1]{#1}

\makeatletter
\newcommand{\aref}[1]{%
\hyperref[#1]{\ref*{#1}(a)}}
\makeatother

\makeatletter
\newcommand{\bref}[1]{%
\hyperref[#1]{\ref*{#1}(b)}}
\makeatother

\begin{document}

\title{Magnetic structure of the magnetoelectric material \bmgo}

\author{A.\,Sazonov}
\email{mail@sazonov.org}
\homepage{http://sazonov.org}
\affiliation{Institute of Crystallography, RWTH Aachen University and J\"ulich Centre for Neutron Science (JCNS) at Heinz Maier-Leibnitz Zentrum (MLZ), 85748 Garching, Germany}
\author{\nc{H.\,Thoma}}
\email{henrik.thoma@frm2.tum.de}
\affiliation{Institute of Crystallography, RWTH Aachen University and J\"ulich Centre for Neutron Science (JCNS) at Heinz Maier-Leibnitz Zentrum (MLZ), 85748 Garching, Germany}
\author{\nc{R.\,Dutta}}
\affiliation{Institute of Crystallography, RWTH Aachen University and J\"ulich Centre for Neutron Science (JCNS) at Heinz Maier-Leibnitz Zentrum (MLZ), 85748 Garching, Germany}
\author{M.\,Meven}
\affiliation{Institute of Crystallography, RWTH Aachen University and J\"ulich Centre for Neutron Science (JCNS) at Heinz Maier-Leibnitz Zentrum (MLZ), 85748 Garching, Germany}
\author{\nc{A.\,Gukasov}}
\affiliation{\nc{Laboratoire L\'{e}on Brillouin, CEA-CNRS, CE-Saclay, 91191 Gif-sur-Yvette, France}}
\author{\nc{R.\,Fittipaldi}}
\affiliation{\nc{CNR-SPIN c/o University of Salerno, 84084 Fisciano, Italy}}
\author{\nc{V.\,Granata}}
\affiliation{\nc{Department of Physics “E.R. Caianiello”, University of Salerno, 84084 Fisciano, Italy}}
\author{T.\,Masuda}
\affiliation{Institute for Solid State Physics, The University of Tokyo, Chiba 277-8581, Japan}
\author{B.\,N\'afr\'adi}
\affiliation{\'Ecole Polytechnique F\'ed\'erale de Lausanne, Laboratory of Nanostructures and Novel Electronic Materials, 1015 Lausanne, Switzerland}
\author{V.\,Hutanu}
\altaffiliation[Present address: ]{FRM II, Technische Universität München, 85747 Garching, Germany}
\email{\\vladimir.hutanu@frm2.tum.de}
\affiliation{Institute of Crystallography, RWTH Aachen University and J\"ulich Centre for Neutron Science (JCNS) at Heinz Maier-Leibnitz Zentrum (MLZ), 85748 Garching, Germany}

\date{\today}
\date{March 1, 2023}


\begin{abstract}
A detailed investigation of \bmgo was performed in its low-temperature magnetoelectric state combining neutron diffraction with magnetization measurements on single crystals. In the paramagnetic state at 10\,K, polarized neutron diffraction was applied to reveal the components of the susceptibility tensor. The crystal and magnetic structures below the antiferromagnetic transition temperature of $T_\text{N} \approx 4$\,K were determined using \nc{unpolarized} neutron diffraction. \nc{This data implies} no structural phase transition from 10\,K down to 2.5\,K and \nc{is} well described within the tetragonal space group $P\bar{4}2_1m$. We found that in zero magnetic field the magnetic space group is either $C_cmc2_1$ or $P_c2_12_12_1$ with antiferromagnetic order along the [110] or [100] directions, respectively, while neighboring spins along the [001] axis are ordered antiferromagnetically. A non-collinear spin arrangement due to small canting within the $ab$ plane is allowed by symmetry and observed experimentally. The ordered moment is found to be 3.24(3)\,$\mu_\text{B}$/Mn$^{2+}$ at 2.5\,K \nc{and the temperature-field dependent magnetic phase diagram is mapped out by macroscopic magnetization}. Distinct differences between the magnetic structure of \bmgo as compared to those of Ba$_2$CoGe$_2$O$_7$ and Ca$_2$CoSi$_2$O$_7$ are discussed.
\end{abstract}


\maketitle


\section{Introduction}


Recently, several members of the melilite family, such as Ca$_2$CoSi$_2$O$_7$, Sr$_2$CoSi$_2$O$_7$, Ba$_2$MnGe$_2$O$_7$ and Ba$_2$CoGe$_2$O$_7$ have been found to exhibit static as well as dynamic magnetoelectric effects~\cite{apl.92.212904.2008,jpcs.150.042001.2009,prb.86.060413.2012,prb.85.174106.2012} and to host remarkable optical properties~\cite{nc.5.1.2014,np.8.734.2012,prl.106.057403.2011}. Moreover, their structural lack of inversion symmetry allows the presence of the antisymmetric Dzyaloshinskii-Moriya interaction~\cite{Dzyaloshinsky1958,Moriya1960,Thoma2022}, which creates in combination with strong spin anisotropies~\cite{Dutta2021,Dutta2023} a rich variety of adopted low-temperature magnetic structures.\cite{Zheludev2003,prb.86.104401.2012,prb.95.174431.2017,Soda2017,prb.81.100402.2010,Zheludev1997} The spin cycloidal \bcugo was even proposed to support a stable skyrmion phase.~\cite{Bogdanov2002}

Due to the lack of the low-temperature structural details, a general non-centrosymmetric tetragonal structure of melilites $A_2BT$O$_7$ with space group (SG) $P\bar{4}2_1m$ is often used.~\cite{nc.5.1.2014,apl.94.212904.2009,prb.89.184419.2014} However, in e.g. Ba$_2$CoGe$_2$O$_7$, the room temperature tetragonal crystal structure $P\bar{4}2_1m$ (Ref.~\onlinecite{jac.49.556.2016}) transforms into orthorhombic $Cmm2$ at low temperatures.~\cite{jac.49.556.2016,prb.89.064403.2014} Another melilite Ca$_2$CoSi$_2$O$_7$ undergoes a series of structural phase transitions from 600\,K down to 30\,K.~\cite{pcm.28.150.2001,acrb.57.271.2001,zfk.215.102.2000,acrb.72.126.2016} To our knowledge, no detailed structural information is available for \bmgo in the magnetic phase.

It is important to note that the main features of the magnetoelectric behaviour of Ba$_2$CoGe$_2$O$_7$ were predicted~\cite{acra.67.264.2011} by symmetry considerations without appealing to any specific atomic mechanism. However, for many melilite systems, specific structural information is still unavailable, while it is clear that the different parent symmetries in the paramagnetic state can easily lead to different ground-state magnetic structures.

It was found that the Mn spins in \bmgo order antiferromagnetically below $T_\text{N}$ with their moments lying in the $ab$ plane.\cite{prb.81.100402.2010,prb.85.174106.2012,Hasegawa2021} The important difference from other melilites, such as \bcgo and \ccso, is that \bmgo is characterised by the non-zero magnetic propagation vector $\bm{k} = (0,0,1/2)$. However, detailed investigation of \bmgo magnetic structure is missing and no information about in-plane spin canting was published. The only neutron diffraction measurements performed on \bmgo consist of only 18 magnetic Bragg reflections\nc{, collected at a cold neutron triple axis spectrometer}.\cite{prb.81.100402.2010}

It is clear that precise information for the atomic positions as well as about the spin order (crystal and magnetic structures) is essential to unravel the complex physics behind the magnetoelectric behavior of the melilite compounds. Recently, we performed~\cite{ic.57.5089.2018} detailed structural investigation on \bmgo by means of neutron diffraction both at room temperature and 10\,K, just above $T_\text{N} \approx 4$\,K. We complemented~\cite{Dutta2022} this study by means of synchrotron X-ray diffraction (XRD), confirming $P\bar{4}2_1m$ as the parent structure with no structural transitions between 110 and 673\,K and revealing the electron density distribution. In order to fill the remaining gap of low temperature crystal (below $T_\text{N}$) and magnetic structure information on \bmgo and to provide reliable data for further experimental and theoretical research, we continued our investigation of \bmgo and performed single-crystal neutron diffraction experiments as well as bulk magnetization measurements at temperatures between 2\,K and 6\,K and fields up to \nc{9}\,T. The observed magnetic properties are compared with those of Ba$_2$CoGe$_2$O$_7$ ($T_\mathrm{N} \approx 6.7$\,K) and Ca$_2$CoSi$_2$O$_7$ ($T_\mathrm{N} \approx 5.7$\,K).


\section{\label{s:exp}Experimental}

Single crystals of \bmgo were grown by floating-zone technique and characterized in previous studies (see Ref.~\onlinecite{acrb.72.126.2016} and references therein). The sample used for the neutron diffraction experiment has a cylindrical shape of approximately 6\,mm in height and about the same diameter. For the magnetization measurements, additional single crystals of \bmgo were provided by the CNR SPIN Salerno. The crystals, grown by floating zone technique, were synthesized starting from polycrystalline rods prepared as described in Ref.~\onlinecite{Granata2017}.

Unpolarized single-crystal neutron diffraction studies were performed on the four-circle diffractometer HEiDi (Ref.~\onlinecite{jlsrf.1.A7.2015}) at the research neutron source FRM\,II at the Heinz Maier-Leibnitz Zentrum (MLZ), Germany. The full data collections for the crystal and magnetic structure refinements were done on HEiDi at 2.5\,K with the wavelength $\lambda = 0.794$\,\AA\ obtained from a Ge(422) monochromator. To reach low temperatures, a closed-cycle He cryostat was mounted in the Eulerian cradle of the diffractometer. The sample was wrapped in Al foil in order to ensure temperature homogeneity. The temperature was measured and controlled by a diode sensor near the heater position. The sample temperature was independently monitored by a second thermometer placed close to the sample position. In addition to the full data collections, temperature dependences of the selected Bragg reflections were measured in the range of 2.5 to 10\,K during the cooling process. The integrated intensities of the measured Bragg reflections were obtained with the DAVINCI program\cite{davinci.sazonov.2015} using the Lehmann-Larsen method for peak location.\cite{acra.30.580.1974} The crystal and magnetic structure parameters of \nc{\bmgo} were refined using the JANA2006 program.\cite{zfk.229.345.2014}

\nc{Additionally, polarized neutron flipping ratios (FRs) were measured on the VIP diffractometer (Ref.~\onlinecite{Gukasov2013}) at the Orph\'{e}e reactor at the Laboratoire L\'{e}on Brillouin (LLB), France. For the measurement at 10\,K in the paramagnetic phase of \bmgo, the single crystal was mounted with the three high symmetry directions [001], [100] and [110] almost along the vertical axis of the instrument and thus parallel to the direction of the applied magnetic field of 6\,T. For each orientation, FR data was collected with a neutron wavelength of 0.84\,\AA{} and refined using the Mag2Pol program.\cite{ Qureshi2019} For the refinement, weak Bragg reflections with intensities below 2\% of the maximal observed value were excluded from the analysis due to a high uncertainty in the calculated FR.}

\nc{Moreover, the field dependence of the magnetization was measured using a vibrating sample magnetometer in a Bitter magnet for fields up to 1.5\,T at the High Field Magnet Laboratory, Nijmegen, and in a superconducting magnet for fields up to 9\,T at the Jülich Center of Neutron science (JCNS), Jülich. The two experiments were performed at $T = 2$ and 1.7\,K with the field parallel to the [110] and the [100], [110] and [001] crystallographic directions, respectively. For the latter experiment, the temperature dependence of the magnetization was additionally measured at different constant field strengths for each of the three high symmetry directions.}


\section{Results and Discussion}

\subsection{Paramagnetic Phase}


\subsubsection{\label{s:paramagnStrModel}Paramagnetic structure model}
\nc{In the paramagnetic (PM) state above $T_N$, the crystal structure of \bmgo is given by the tetragonal space group $P\bar{4}2_1m$ with structural details as reported in Ref.~\onlinecite{ic.57.5089.2018}. For applied magnetic fields, an induced ferromagnetic (FM) moment at the Mn ions is expected. The relation of this FM moment $\vb{m}^{\text{FM}}$ to the external magnetic field $\vb{B}$ can be described in terms of a susceptibility tensor $\chi$, resulting in $\vb{m}^{\text{FM}}=\chi\vb{B}$. For small fields up to around 6\,T, the values of the symmetric tensor $\chi$ are expected to be almost constant in the PM phase at 10\,K,\cite{prb.85.174106.2012} but can be anisotropic with components restricted to $\chi_{11}=\chi_{22}$ and $\chi_{12}=\chi_{13}=\chi_{23}=0$ by the tetragonal symmetry. However, macroscopic measurements of the low temperature magnetic susceptibility and the room-temperature electron spin resonance for \bmgo revealed only an almost negligible anisotropy, which can be related to the zero orbital angular momenta of Mn$^{2+}$ in the high spin state.\cite{prb.81.100402.2010}}


\subsubsection{\label{s:paramagnStrRef}Field-induced magnetic structure at 10\texorpdfstring{\,}{\xspace}K}

\begin{table}[b]
\caption{\label{t:ParamagnStr}
Field-induced magnetic moment of the Mn atoms in the paramagnetic phase of \bmgo, as refined from polarized neutron diffraction data at 10\,K with different applied magnetic field directions. The number of measured reflections for each FR dataset and the reduced refinement residuals $\chi^2_r$ are additionally listed. The refinement is based on the structural parameters given in Ref.~\onlinecite{ic.57.5089.2018}.}
\begin{ruledtabular}\vspace{1ex}
\begin{tabular}{lccc}
Magnetic field & Total reflections &  $m^{\text{FM}}\ [\mu_\text{B}]$   & $\chi^2_r$      \\
\colrule
$B\parallel[001]$, 6\,T    & 630 &  1.912(8)    & 3.37  \\
$B\parallel[100]$, 6\,T   & 724  &  1.957(3)    & 3.64  \\
$B\parallel[110]$, 6\,T   & 775  &  2.005(3)    & 2.81  \\
\end{tabular}
\end{ruledtabular}
\end{table}

\begin{figure}[t]
\includegraphics[width=1\columnwidth]{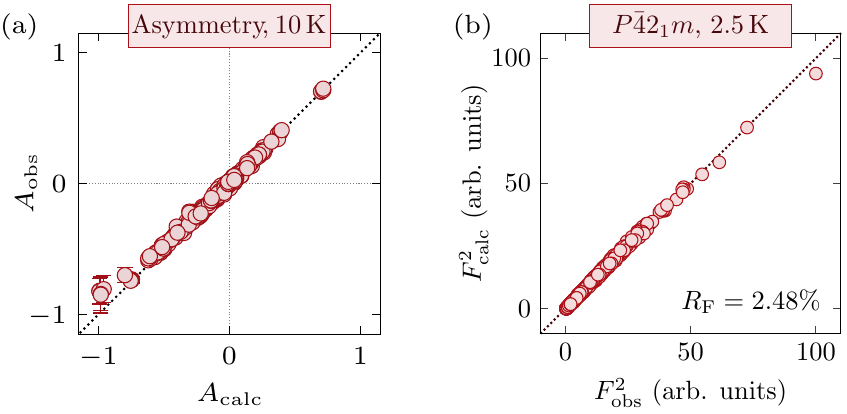} %
\caption{\label{f:fitQuality}(Color online) (a) Observed and calculated asymmetry values for all $h+k$ even reflections measured in \bmgo at $10$\,K with different applied magnetic field directions. The calculation is based on the refinement results in Tab.~\ref{t:ParamagnStr}. (b) Quality of the \bmgo crystal structure refinement in SG $P\bar{4}2_1m$ according to presented single-crystal neutron diffraction data at 2.5\,K. Experimentally measured integrated intensities ($F^2_\mathrm{obs}$) are plotted against the calculated ones ($F^2_\mathrm{calc}$). Reliability factor $R_\text{F}$ is also given.}
\end{figure}

\nc{To determine the local susceptibility tensor and to probe the anisotropy in the field-induced magnetization, $\vb{m}^{\text{FM}}$ is refined from the polarized neutron FR data, collected at 10\,K in the PM phase of \bmgo with magnetic field applied along different crystallographic directions. For the refinement, the structure parameters were fixed according to the results of the unpolarized neutron diffraction, presented in Ref.~\onlinecite{ic.57.5089.2018}. The refined values of the field induced magnetization $m^{\text{FM}}$, directed along the applied field magnetic field, are listed in Table~\ref{t:ParamagnStr}. The very good agreement between the measured and calculated asymmetry values~\footnote{The refinement quality of $m$ parameters from the polarized neutron diffraction data of $n$ reflections is given by the reduced chi-square value of $\chi^2_r = 1/(n-m)\sum (A_\mathrm{obs} -  A_\mathrm{calc})^2/dA_\mathrm{obs}^2$. The asymmetry $A$ is related to the classical FR value $R$ by $A=(R-1)/(R+1)$} is demonstrated in Fig.~\aref{f:fitQuality} for all reflections that are sensitive to $m^{\text{FM}}$, i.e.\ reflections with $h+k$ even. Note that reflections with $h+k$ odd are only sensitive to Mn moments that are antiferromagnetically ordered within the ab plane; thus, their asymmetry is zero in the PM phase.}

\nc{For all three high-symmetry field directions listed in Table~\ref{t:ParamagnStr}, a similar induced magnetic moment is refined. Based on a least-squares refinement of these these FM moments taking into account the exact field direction from the experimental orientation matrix, almost equal susceptibility components of $\chi_{11}=0.330(4)$ and $\chi_{33}=0.32(2)$ are determined. They indicate a negligible anisotropy of $\chi_{11}/\chi_{33}=1.04(5)$, which is in agreement with the expectation from $g$ tensor measurements by electron spin resonance.\cite{prb.81.100402.2010} The average induced magnetic moment of of $1.96(4)\mu_B/$Mn for an applied field of $6$\,T at $10$\,K in \bmgo is close to the value obtained by macroscopic magnetization.\cite{prb.85.174106.2012}  Note that no polarized neutron diffraction data was collected in \bmgo below \TN as the ordered antiferromagnetic (AFM) moments cannot be probed by the FR method due to $\vb{k} = (0,0,1/2)$.}


\subsection{Antiferromagnetic Phase}


\subsubsection{\label{s:crystStr}Crystal structure details at 2.5\texorpdfstring{\,}{\xspace}K}

In order to determine the structural parameters for \bmgo below the magnetic transition temperature $T_\text{N} \approx 4$\,K we performed a refinement of its crystal structure from our \nc{unpolarized} neutron diffraction data at 2.5\,K. A total of 1657 pure nuclear Bragg reflections with $\ensuremath{\sin\theta/\lambda} \lesssim 0.8$\,\AA$^{-1}$ were measured and 964 unique reflections were obtained by averaging equivalents~\footnote{The accuracy of the averaged intensities is estimated from the spread of the individual measurements of equivalent reflections by $R_\mathrm{int} = \sum|F^2_\mathrm{obs} - \protect\langle F^2_\mathrm{obs}\protect\rangle| / \sum F^2_\mathrm{obs}$.} ($R_\mathrm{int} = 0.0223$).

No indication of a symmetry change was found from 10\,K down to 2.5\,K according to the neutron diffraction data. Thus, within the experimental precision, it is assumed that the crystal structure of \bmgo in the magnetic phase coincides with that at 10\,K (tetragonal SG \PFTOm), as determined by our previous neutron diffraction studies.\cite{ic.57.5089.2018} These structural details were taken as starting parameters for the refinement at 2.5\,K. All atomic positions which are not restricted by symmetry were refined together with the scale, extinction and anisotropic atomic displacements ($U_\mathrm{ani}$) parameters. Besides this, no other constraints were used in the refinement process.

The agreement between the experimental and calculated data is shown in Fig.~\bref{f:fitQuality}. Table~\ref{f:CrystStrParams} presents the refined atomic coordinates as well as isotropic atomic displacement ($U_\mathrm{iso}$) parameters. Full details of the refinement, including $U_\mathrm{ani}$ parameters, bond lengths and angles, are deposited in the crystallographic information file (CIF).\cite{icsd.ccso.2017} A comparison with the 10\,K structure from our neutron diffraction~\cite{ic.57.5089.2018} shows negligible differences in positional parameters with an average value of less than $1\sigma$.


\begin{table}
\caption{\label{f:CrystStrParams}Fractional atomic coordinates $x$, $y$ and $z$ as well as isotropic atomic displacement parameters $U_\mathrm{iso}$ (\AA$^2$) for \bmgo refined in SG $P\bar{4}2_1m1'$ according to present single-crystal neutron diffraction data at 2.5\,K.}
\begin{ruledtabular}
\begin{tabular}{llllll}
Ion   & WP  & $x$         & $y$         & $z$          & $U_\mathrm{iso}$ \\
\colrule
Ba    & 4e  & 0.33475(7)  & $0.5-x$     & 0.49278(14)  & 0.00243(14)      \\
Mn    & 2b  & 0           & 0           & 0            & 0.00290(30)      \\
Ge    & 4e  & 0.13902(4)  & $0.5-x$     & 0.04574(8)   & 0.00240(9)       \\
O1    & 2c  & 0           & 0.5         & 0.16249(18)  & 0.00410(20)      \\
O2    & 4e  & 0.13834(7)  & $0.5-x$     & 0.73340(13)  & 0.00547(14)      \\
O3    & 8f  & 0.07813(7)  & 0.18921(7)  & 0.19522(10)  & 0.00504(13)      \\
\end{tabular}
\end{ruledtabular}
\end{table}


\begin{figure*}[t]
\includegraphics[width=1.5\columnwidth]{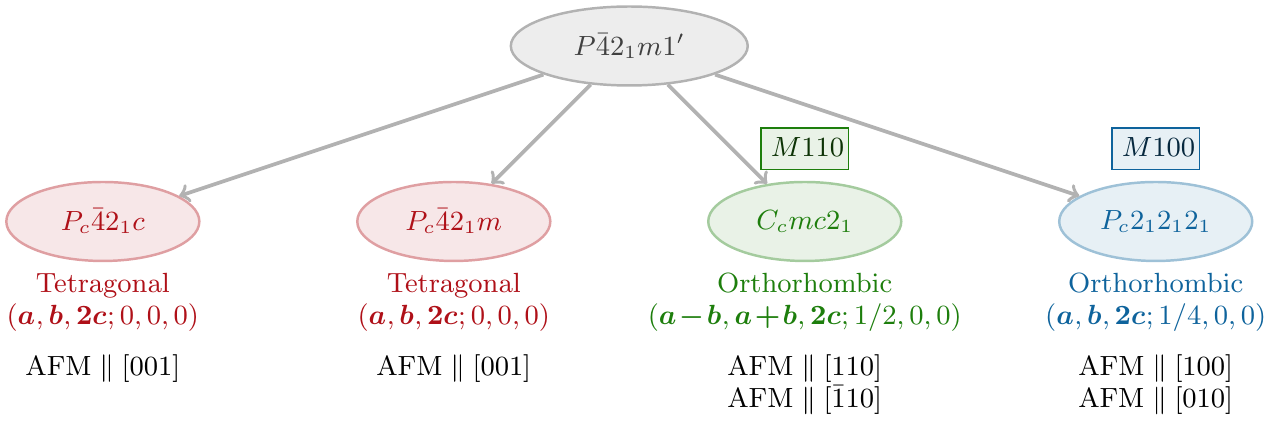}
\caption{\label{f:MsgDiag}(Color online) The possible $k$-maximal symmetries for a magnetic ordering of Mn with propagation vector $\bm{k} = (0,0,1/2)$ on a paramagnetic phase with SG $P\bar{4}2_1m$ and its corresponding gray group $P\bar{4}2_1m1'$. Only the subgroups which allow nonzero magnetic moments are shown. Each magnetic space group label is shown together with the transformation from the parent $P\bar{4}2_1m$ unit cell basis $\{\bm{a}, \bm{b}, \bm{c}\}$ to its standard setting. The allowed magnetic moment components which correspond to the antiferromagnetic (AFM) ordering are given at the bottom.}
\end{figure*}

\subsubsection{\label{s:magnStrModel}Antiferromagnetic structure models}

The low-temperature crystal structure of \bmgo is well described by the tetragonal space group \PFTOm (see Sec.~\ref{s:crystStr} for more details). Below $T_\text{N} \approx 4$\,K, additional Bragg reflections appear corresponding to the magnetic order with propagation vector $\bm{k} = (0,0,1/2)$. Thus, the unit cell of the magnetic structure is doubled along the $c$ axis compared to the paramagnetic state.

In order to solve the magnetic structure of \bmgo we used the concept of Shubnikov groups (magnetic space groups, MSGs), which is very useful in the case of second-order phase transitions for enumerating the possible magnetic structures compatible with the crystal symmetry. This approach implies specific symmetry-deduced constraints on the magnetic moments. That is, the magnetic moments of symmetry equivalent atoms are related via the magnetic symmetry operations. It allows to reduce the number of refined parameters and to average the symmetry equivalent reflections. It was shown, that the use of MSGs significantly facilitates the interpretation of the results (see, e.g., Ref.~\onlinecite{armr.45.217.2015} and references therein).

The non-magnetic parent space group of \bmgo is $P\bar{4}2_1m$ and its corresponding gray group is $P\bar{4}2_1m1'$, which in addition includes the time reversal operation. The symmetry of a magnetically ordered phase is described by a subgroup of this parent group. Fig.~\ref{f:MsgDiag} shows the $k$-maximal subgroups for $P\bar{4}2_1m1'$ with the magnetic propagation vector $\bm{k} = (0,0,1/2)$. Only four distinct types of magnetic ordering of $k$-maximal symmetry are possible in \bmgo: MSG $P_c\bar{4}2_1c$, MSG $P_c\bar{4}2_1m$, MSG $C_cmc2_1$ (model $M110$) and MSG $P_c2_12_12_1$ (model $M100$). The differences between the models manifest themselves in the distinct magnetic modes which can be presented as the pure antiferromagnetic (AFM) components along the crystallographic directions [001], [110] or [100] (see Fig.~\ref{f:MsgDiag}).

\begin{figure}[t]
\includegraphics[width=1\columnwidth]{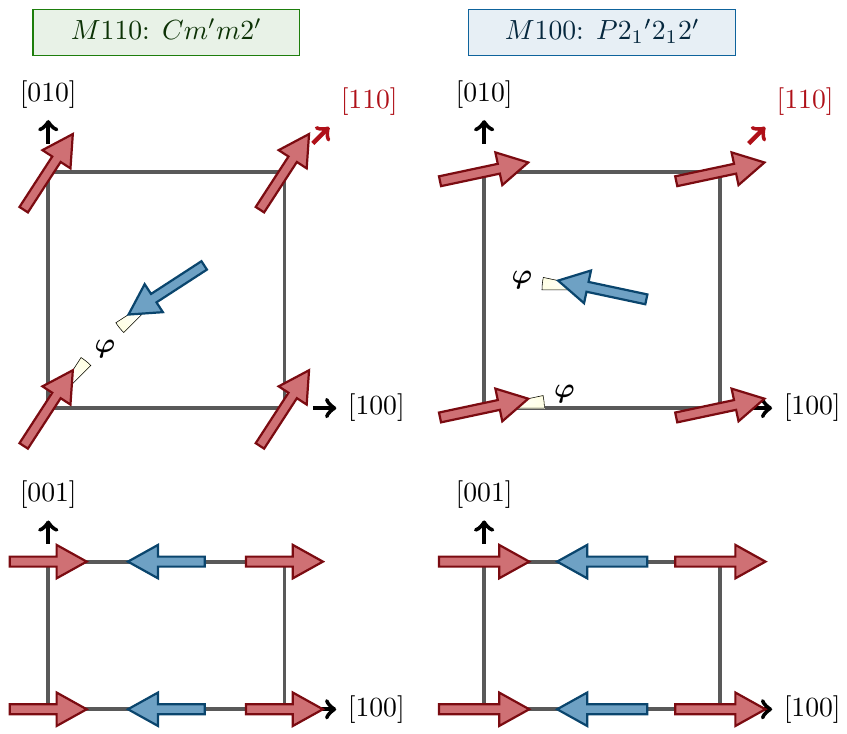}
\caption{\label{f:MagnStr}(Color online) Possible in-plane magnetic structures of \bmgo in the $M110$ (MSG $C_cmc2_1$) and  $M100$ (MSG $P_c2_12_12_1$) models, according to our neutron diffraction data at 2.5\,K. Angle $\varphi$ denotes in-plane canting of magnetic moments allowed by symmetry.}
\end{figure}

In the case of tetragonal magnetic models (MSGs $P_c\bar{4}2_1c$ and $P_c\bar{4}2_1m$) the magnetic moment is aligned along the $c$ axis. That disagrees with the experimentally observed magnetic Bragg peaks. Therefore, these two models can be ruled out. The orthorhombic magnetic models $M110$ and $M100$ (see Fig.~\ref{f:MagnStr}) restrict the magnetic moment to lie in the $ab$ plane with a main AFM component along either [110] diagonal or [100] axis. A minor AFM component in the perpendicular direction (in-plane canting of magnetic moments) is allowed by symmetry, but does not lead to a global ferromagnetic moment because of the AFM alignment between the (001) planes. Therefore, in \bmgo the FM alignment is forbidden by symmetry in contrast to other compounds in the melilite family, such as \bcgo or \ccso.

\begin{figure}[t]
\includegraphics[width=1\columnwidth]{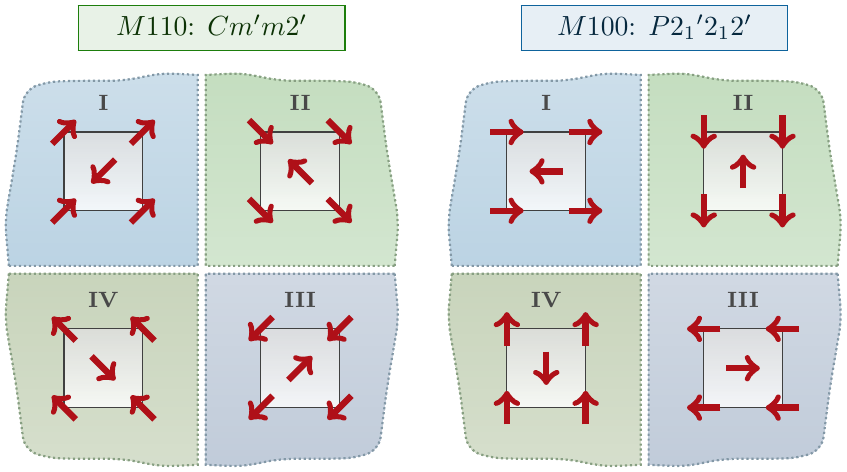}
\caption{\label{f:MagnDomains}(Color online) Schematic view of the possible magnetic domains in \bmgo for both $M110$ (MSG $C_cmc2_1$) and  $M100$ (MSG $P_c2_12_12_1$) models. A single layer along the [001] direction is shown for simplicity.}
\end{figure}

In both $M110$ and $M100$ the index of their MSGs with respect to the parent symmetry $P\bar{4}2_1m1'$ is four. Thus, two configurations twinned with respect to a diagonal mirror plane $m_{xy}$ ($90^\circ$ magnetic domains) exist in \bmgo, apart from their corresponding trivial twins with all spins reversed. All possible magnetic domains for both models are shown schematically in Fig.~\ref{f:MagnDomains}.

\subsubsection{\label{s:magnStrRef}Magnetic structure refinement at 2.5\texorpdfstring{\,}{\xspace}K}

Two symmetry-allowed magnetic structure models ($M110$ and $M100$) selected in the previous section were used in the refinement process to compare the experimental data with the calculations. We note, that in both cases the 90\degree magnetic domains were taken into account in the refinement, while the trivial magnetic twins with all spins reversed were omitted, since they have no effect on the diffraction data. We note, however, that if the 90\degree magnetic domains are equally populated, it becomes impossible to distinguish between $M110$ and $M100$ with unpolarized neutron diffraction only.

In contrast to \bcgo (Refs.~\onlinecite{prb.86.104401.2012,prb.89.064403.2014}) or \ccso (Ref.~\onlinecite{prb.95.174431.2017}), \bmgo is characterised by the non-zero magnetic propagation vector $\bm{k} = (0,0,1/2)$. As a result, the magnetic Bragg reflections are fully separated from the nuclear ones. Therefore, in the refinement, all structural parameters for \bmgo were fixed according to the results of the crystal structure analysis presented in Sec.~\ref{s:crystStr} and only the magnetic structure components were refined. A total of 181 pure magnetic Bragg reflections with $\ensuremath{\sin\theta/\lambda} \lesssim 0.46$\,\AA$^{-1}$ were measured at 2.5\,K and 91 unique reflections were obtained by averaging equivalents~\footnote{The accuracy of the averaged intensities is estimated from the spread of the individual measurements of equivalent reflections by $R_\mathrm{int} = \sum|F^2_\mathrm{obs} - \protect\langle F^2_\mathrm{obs}\protect\rangle| / \sum F^2_\mathrm{obs}$.} ($R_\mathrm{int} = 0.0275$).

\begin{figure}[t]
\includegraphics[width=1\columnwidth]{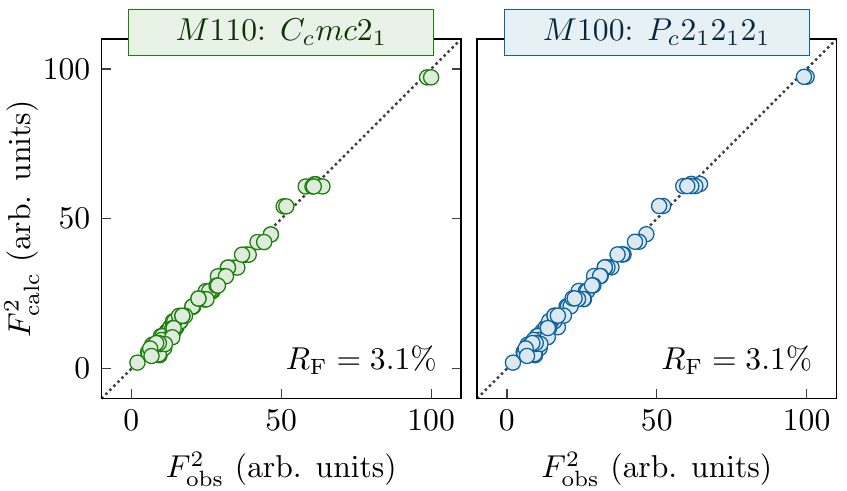}
\caption{\label{f:MagnStrFit}(Color online) Quality of the \bmgo magnetic structure refinement in both $M110$ (MSG $C_cmc2_1$) and $M100$ (MSG $P_c2_12_12_1$) models according to present syngle-crystal neutron diffraction data refinement at 2.5\,K. Experimentally measured integrated intensities ($F^2_\mathrm{obs}$) are plotted against the calculated ones ($F^2_\mathrm{calc}$). Reliability factors $R_\text{F}$ are also given.}
\end{figure}

A direct comparison of the entire set of pure magnetic reflections for the final fit in the two models $M110$ and $M100$ is presented in Fig.~\ref{f:MagnStrFit}. As can be seen from the figure, both models reproduce the magnetic reflection intensities equally well, and it is impossible to distinguish between the two cases. This indicates that the magnetic domains are equally populated or their imbalance is negligible and cannot be easily detected. The average ordered magnetic moment in \bmgo at 2.5\,K, as refined from neutron diffraction data, is found to be about 3.24(3)\,$\mu_\mathrm{B}$/Mn for both $M110$ and $M100$ models (see Table~\ref{t:MagnStr}). In both cases, the spin canting is allowed by the symmetry via the minor AFM components perpendicular to the direction of the primary AFM ordering. This minor AFM moment is found to be around 0.4\,$\mu_\mathrm{B}$, which is close to the limit of unpolarised neutron diffraction.

\begin{table}[b]
\newcommand\Mx{\multicolumn{1}{c}{$M_x$}}
\newcommand\My{\multicolumn{1}{c}{$M_y$}}
\newcommand\Mz{\multicolumn{1}{l}{\,$M_z$}}
\newcommand\MM{\multicolumn{1}{c}{$|M|$}}
\caption{\label{t:MagnStr}
Magnetic moment components ($\mu_\mathrm{B}$) of the symmetry-independent Mn atoms in the $M110$ (MSG $C_cmc2_1$) and $M100$ (MSG $P_c2_12_12_1$) magnetic structures models, as refined from neutron diffraction data at 2.5\,K. Manganese atomic coordinates are given in Table~\ref{f:CrystStrParams}.}
\begin{ruledtabular}\vspace{1ex}
\begin{tabular}{lcccc}
Model     & \Mx       & \My      & \Mz   & \MM      \\
\colrule
$M110$    &  2.57(4)  & 1.97(4)  &  0    & 3.24(3)  \\
$M100$    & -0.43(5)  & 3.21(3)  &  0    & 3.24(3)  \\
\end{tabular}
\end{ruledtabular}
\end{table}


\subsubsection{\label{s:magnStrEvolT}Temperature evolution of magnetic structure}

\begin{figure}[t]
\includegraphics[width=1\columnwidth]{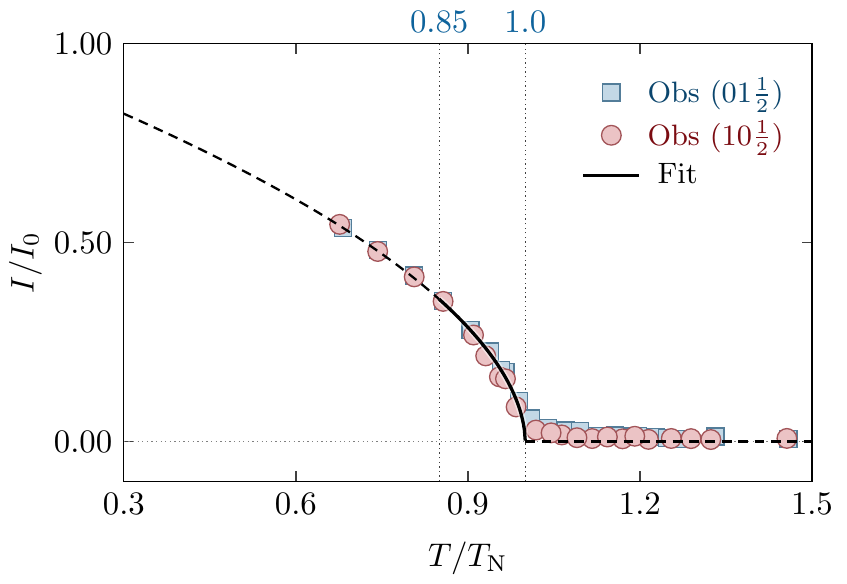}
\caption{\label{f:IvsT}(Color online) Temperature dependences of the normalised integrated intensity of the magnetic Bragg reflections (10$\frac{1}{2}$) and (01$\frac{1}{2}$). The symbols correspond to the single-crystal neutron diffraction data with error bars too small to be visible. The solid line shows a fit to Eq.~\ref{eq:IvsT}.}
\end{figure}

In order to follow the temperature evolution of the magnetic structure of \bmgo, several intense magnetic and structural Bragg reflections were collected in the temperature range from 2.5\,K to 6\,K. Figure~\ref{f:IvsT} shows the temperature dependences of the normalised integrated intensities of the magnetic Bragg reflections (10$\frac{1}{2}$) and (01$\frac{1}{2}$), as an example. These intensities decrease continuously with increasing temperature and become constant and close to zero above $T_{\rm N}$.

The integrated intensities $I$ of magnetic Bragg reflections measured with unpolarized neutrons follow the square of the magnetic order parameter. The data were fitted in the temperature range from 0.8$T_{\rm N}$ to $T_{\rm N}$ assuming a power law dependence to the equation\cite{book.chatterji.2006,prb.83.134438.2011}

\begin{eqnarray}
\label{eq:IvsT}
I = I_{\rm n} + I_0 \left( \frac{T_{\rm N} - T}{T_{\rm N}} \right) ^ {2\beta},
\end{eqnarray}
where $I_{\rm n}$ is the nuclear (structural) contribution to the intensity, $I_0$ is the magnetic intensity at $T=0$ and $\beta$ is the critical exponent.

The fit yields $\beta = 0.27 \pm 0.05$ as critical exponent which is close to the values found for Ba$_2$CoGe$_2$O$_7$ (Ref.~\onlinecite{prb.89.064403.2014}), \ccso (Ref.~\onlinecite{prb.95.174431.2017}) and other layered 2D antiferromagnets.~\cite{prl.72.1096.1994,prb.49.8811.1994,zpb.96.465.1995} This is in agreement with the layered crystal structure of \bmgo where MnO$_4$ and Ge$_2$O$_7$ groups in the planes are separated by interlayer Ba cations and exchange interactions between manganese spins on neighboring layers are expected to be much weaker than intra-plane exchange couplings.

A very weak nuclear contribution above $T_{\rm N}$ is associated with the multiple scattering, also known as Renninger effect (see, e.g., Refs.~\onlinecite{zp.106.141.1937}--\onlinecite{acra.62.174.2006} and references therein), which was found in Ca$_2$CoSi$_2$O$_7$ at 10\,K (Ref.~\onlinecite{acrb.72.126.2016}) and Ba$_2$CoGe$_2$O$_7$ at room temperature (Ref.~\onlinecite{jac.49.556.2016}).


\subsubsection{\label{s:magnStrEvolH}Magnetization measurements}

\begin{figure}[t]
\includegraphics[width=1\columnwidth]{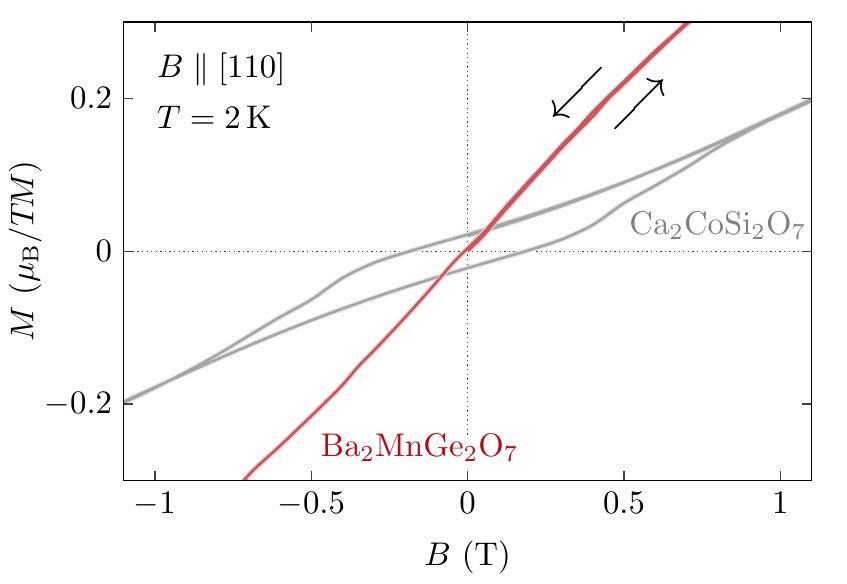}
\caption{\label{f:MvsHlow}(Color online) Field dependence of the magnetization $M$ in $\mu_\text{B}$ per transition metal ion (TM) for \ccso (blue) and \bmgo (red). Field is applied along the [110] direction at 2\,K. In contrast to \ccso, there is no hysteresis for \bmgo indicating no spontaneous magnetization in the $ab$ plane in agreement with the magnetic symmetry analysis.}
\end{figure}

\begin{figure}[b]
\includegraphics[width=1\columnwidth]{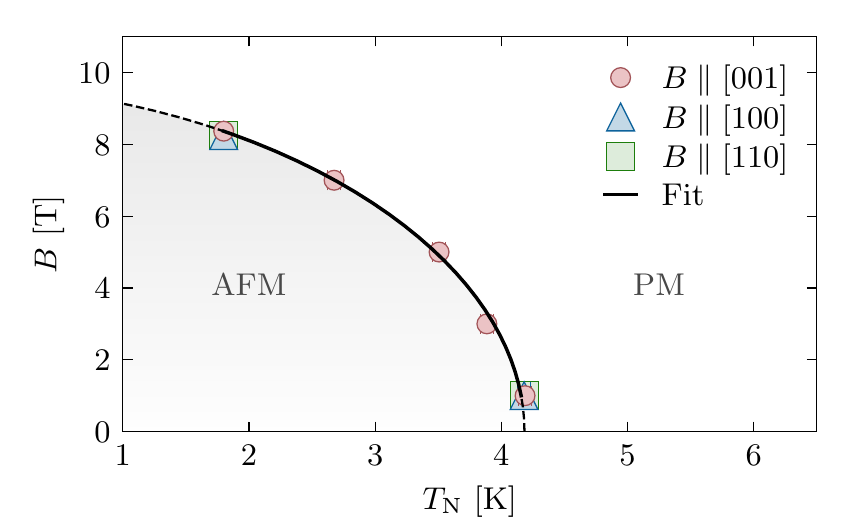}
\nc{\caption{\label{f:bcumngo_mt_3N_phases}(Color online)  Field- and temperature-dependent magnetic phase diagram in \bmgo. Symbols correspond to the magnetization measurements with field applied in the respective crystallographic direction. The solid line shows a fit of the $B\parallel[001]$ data (red circular symbols) to Eq.~\ref{eq:TNvsB}.}}
\end{figure}

The absence of a global ferromagnetic component derived from the magnetic symmetry analysis (see Sec.~\ref{s:magnStrModel}) agrees well with the absence of in-plane spontaneous magnetization observed in \bmgo below $T_\text{N} \approx 4$\,K. Figure~\ref{f:MvsHlow} presents the results of the bulk magnetization ($M$) measurements at 2\,K, just below the N\'eel temperature, with field applied along the [110] direction. For comparison, the magnetization curve of another member of the melilite family, \ccso, shows a clear spontaneous magnetization, both observed experimentally and allowed by symmetry (Ref.~\onlinecite{prb.95.174431.2017}). Unfortunately, the absence of in-plane spontaneous magnetization makes it impossible to independently estimate the mean value of zero field canting within the $ab$ plane, as was successfully done earlier for \bcgo (Ref.~\onlinecite{prb.95.174431.2017}).

\nc{In addition, field- and temperature-dependent magnetization measurements were used to outline the magnetic phase diagram by monitoring the change in the transition temperature \TN for increasing external fields $B$. The results are shown in Fig.~\ref{f:bcumngo_mt_3N_phases} and clearly indicate a decrease in \TN for higher fields, independent on the applied field direction. This isotropic behaviour complies well with the polarized neutron diffraction results in Sec.~\ref{s:paramagnStrRef}. The reduction of \TN for stronger external fields is typical for antiferromagnets and connected to the field-induced suppression of the order parameter. Applying the Ising model of antiferromagnets, the general variation of \TN with applied magnetic fields $B$ was derived by \citet{Bienenstock1966} using a high-temperature expansion of the free energy and reduced to an expression of the form
\begin{equation}
\frac{T_N}{T_{N,0}} = \left(1-\left(\frac{B}{B_0}\right)^2\right)^{\zeta},
\label{eq:TNvsB}
\end{equation}
with zero-field transition temperature $T_{N,0}$, zero-temperature critical field $B_0$ and exponent $\zeta$. Note that the critical field $B_0=-zJ/m$ is given by the coordination number $z$, the spin coupling energy $J$ and the magnetic moment $m$. Using this approximation, the field dependency can be well fitted by $T_{N,0}=4.18(4)$\,K, $B_0=9.5(3)$\,T and $\zeta=0.58(8)$, as shown by the solid black line in Fig.~\ref{f:bcumngo_mt_3N_phases}. Its continuation, given by the dashed line, serves as good approximation for the boundary between the AFM and PM phase. As the exponent $\zeta$ reflects mainly the dimension of the system and is around $0.87$ for a 2D square lattice and around $0.35$ for a simple or body centered cubic 3D structure\cite{Bienenstock1966}, the fitted result of $0.58(8)$ indicates the quasi-2D AFM character of \bmgo, resulting from its layered structure discussed in Sec.~\ref{s:magnStrEvolT}.}


\section{Conclusion}

In the paramagnetic phase of \bmgo, the field induced magnetization was precisely refined from polarized neutron diffraction measurements. The results indicate no clear anisotropy in the local susceptibility tensor, which is in agreement with the expectation from electron spin resonance~\cite{prb.81.100402.2010} and our magnetization measurements.

For the low-temperature magnetoelectric state, high-quality structural parameters of \bmgo are reported. No evidence for a structural phase transition upon the magnetic phase transition at about 4\,K was observed by neutron diffraction and the crystal structure at 2.5\,K is found to correspond well to that at 10\,K.

\nc{In addition, the} magnetic structure of \bmgo at 2.5\,K was accurately refined based on the single-crystal neutron diffraction data and magnetic symmetry analysis. The results indicate the orthorhombic symmetry of the magnetic structure with either $C_cmc2_1$ or $P_c2_12_12_1$ magnetic space group, which is impossible to distinguish with unpolarized neutron diffraction because of the equally populated 90$^\circ$-type magnetic domains. In both models, the spin pattern shows a square-lattice in-plane AFM order along the $[110]$ (MSG $C_cmc2_1$) or $[100]$ (MSG $P_c2_12_12_1$) crystallographic directions. At zero magnetic field the magnitude of the averaged ordered magnetic moment of Mn$^{2+}$ ions is found to be about 3.24\,$\mu_\text{B}$. Small canting (minor antiferromagnetic component) in the $ab$ plane perpendicular to the primary AFM moment is allowed by symmetry and found to be relatively small.

The detailed structural parameters (both for the nuclear and magnetic order) \nc{and the magnetic phase diagram} of \bmgo reported here can serve as a profound experimental basis to develop microscopic models describing the multiferroic nature and the peculiar magnetoelectric phenomena in melilites. 
\nc{Uniaxial or spherical neutron polarization analysis} on the sample with an imbalance of the 90\degree magnetic domains would help to distinguish between two magnetic models presented here and thus provide further insights into the nature of the magnetic phases of Co- and Mn-based melilites.


\begin{acknowledgments}
This work was partly supported by the BMBF under contract No.\,05K13PA3. Part of the work is based upon experiments performed at the HEiDi instrument which is operated by RWTH Aachen University and Forschungszentrum J\"ulich GmbH (J\"ulich Aachen Research Alliance JARA).
\end{acknowledgments}


\bibliography{manuscript}
\end{document}